\documentclass[prb,showpacs,showkeys,preprintnumbers,amsmath,amssymb,twocolumn]{revtex4-1}
\usepackage[T1]{fontenc} 
\usepackage{makeidx} 
\usepackage{graphicx} 
\usepackage{dcolumn} 
\usepackage{array} 
\usepackage{amssymb} 
\usepackage{amsmath}
\usepackage{textcomp}
\usepackage{rotating}
\usepackage{wasysym}
\usepackage{multirow}
\usepackage{subfigure}
\usepackage{eucal}
\usepackage{mathrsfs}
\usepackage{units}
\usepackage{rotating}
\usepackage[all]{xy}
\usepackage{float} 
\usepackage{amsmath} 
\usepackage{amsfonts} 
\usepackage{bm} 
\usepackage[amssymb]{SIunits}

\begin{document}

\title{Influence of longitudinal electric field on the oscillating magnetocaloric effect of graphenes}

\author{M.S. Reis}\email{marior@if.uff.br}\affiliation{Instituto de Física, Universidade Federal Fluminense, Av. Gal. Milton Tavares de Souza s/n, 24210-346, Niterói-RJ, Brasil}

\date{\today}

\begin{abstract}
The present effort explores the influence of a longitudinal applied electric field on the magnetocaloric properties of graphenes. The magnetic entropy change $\Delta S(T,\Delta B,E)$ has two contributions: $S_{cos}(T,B,E)$ with an oscillating character on $m$, inversely proportional to the magnetic field; and $S_{per}(T,B,E)$ with a periodic character also on $m$. In comparison to the case without electric field, the maximum value of the magnetic entropy change either increases or decreases due to an applied electric field, depending on the value of $m$; and, in addition, the temperature in which the maximum entropy change occurs decreases due to the electric field.
\end{abstract}

\maketitle

The magnetocaloric effect (MCE) is an intrinsic and exciting property of magnetic materials, in which, under an isothermal process, they are able to exchange heat ($\Delta Q=T\Delta S$) with a thermal reservoir, under a magnetic field change $\Delta B$. This amount of heat is measured from the magnetic entropy change $\Delta S(T,\Delta B)$, usually used to characterize materials concerning the magnetocaloric properties. From the adiabatic process, under a magnetic field change, the magnetic material changes temperature.


In addition, the MCE has a significant technological
importance, since magnetic materials with large MCE values can
be employed in various thermal devices\cite{tishin_book} - promising candidates to substitute, in a near feature, the common devices, i.e., those based on compression-expansion of Freon-like gases (for instance, air conditioners and household refrigerators). This probable scenario arises since magnetic cooling machines fit in a clean, safe and sustainable technology; where these systems do not use any noxious gases to the ozone layer and, in addition, have greater efficiency of cooling power and lower energy consumption when compared with the traditional devices. However, the magnetocaloric potential of materials is not limited to room temperature applications; and the best example is the adiabatic demagnetization refrigerator, that can reach mK scale \cite{Timbie1990271}.

In spite of those direct applications, the inverse magnetocaloric effect (negative magnetic entropy change), is opening doors for devices, since the reverse operation can be done (cooling instead of heating, when the magnetic material is submitted to a magnetic field change under adiabatic conditions). In this sense, a lot of efforts have been done by the community \cite{ISI:000262375100025,ISI:000276275300037,ISI:000277373400002,ISI:000278911500037,ISI:000281216800018,ISI:000288569300041,ISI:000289373000001}.

Nowadays, the magnetic materials available and studied by the scientific community do not have yet the needed characteristics to be used in large scale, due to technological and/or economical restrictions; and the literature on the magnetocaloric effect sum efforts for applications considering only materials with cooperative orderings, like ferro-, antiferro- and even ferrimagnetic materials\cite{tishin_book}. However, recently, diamagnetic materials received due attention never received before \cite{MCE_OSC_DT, MCE_OSC_DS}, and an oscillatory behavior (named oscillating magnetocaloric effect - OMCE), due to the crossing of the Landau levels through the Fermi level \cite{greiner}, was predicted, in analogy to the de Haas-van Alphen effect.

To optimize the effect, other diamagnetic materials have to be considered; however, 2D standard systems have a energy spectra similar to the 3D case and therefore a 2D massless Dirac systems, more precisely graphenes, were the natural choice\cite{mce_grafeno}. 

Graphene is a planar sheet of Carbon atoms packed in a honeycomb lattice. Electronic properties of graphene are quite different from standard 3D diamagnetic materials, since electrons in graphene have zero effective mass and these behave like relativistic particles, described by the Dirac equation\cite{livro_grafeno}. While for a non-relativistic 3D diamagnetic material the Landau levels are given by a harmonic oscillator\cite{reis_book,livro_grafeno}:
\begin{equation}
E_i=\hbar\omega\left(i+\frac{1}{2}\right)\;\;\;\;\;\;\textnormal{where}\;\;\;\;\;\;\hbar\omega=\frac{\hbar eB}{m}
\end{equation}
for a 2D massless Dirac material the levels are given by\cite{livro_grafeno}:
\begin{equation}
E_j=\hbar\omega^\prime\sqrt{j}\;\;\;\;\;\;\textnormal{where}\;\;\;\;\;\;\hbar\omega^\prime=\sqrt{2\hbar eB}v_F
\end{equation}
where $i,j=0,1,2,\cdots$  represents the Landau levels and $v_F=10^6$ m/s stands for the Fermi velocity (only 300 times smaller than the speed of light). Note the energy spectra for a non-relativistic material are equally spaced, while it is not for a 2D massless Dirac material.

A 3D non-relativistic diamagnetic materials (like Gold), has the OMCE maximized at low temperatures (c.a. 1 K)\cite{MCE_OSC_DS, MCE_OSC_DT}; while the maximum OMCE of a 2D massless Dirac system (a graphene), appears in a more comfortable temperature range (c.a. 100 K)\cite{mce_grafeno}, due to the relativistic properties of the electrons into this material, more precisely the huge Fermi velocity. In addition, for a 3D non-relativistic material, the change of magnetic field change needed to invert the magnetocaloric effect from normal to inverse is c.a. 1 mT\cite{MCE_OSC_DS, MCE_OSC_DT} (useful for application in high sensible magnetic field sensor at low temperatures); while for a graphene it is c.a. 3.4 T\cite{mce_grafeno}. Following these results, the aim of the present paper is to understand the influence of a longitudinal electric field on the OMCE of graphenes, in order to further optimize the effect.

The energy spectra for a nano-ribbon graphene sheet on $x-y$ plan with $\vec{E}=(-E,0,0)$ and $\vec{B}=(0,0,B)$, i.e., electric field along the graphene plan and magnetic field perpendicular to graphene plan, is\cite{JPCM_22_2010_115302, PLA_375_2011_3624,prl_referee}:
\begin{equation}
E_n=\hbar\omega^{\prime\prime}\sqrt{n}+\hbar v_F\beta k_y
\end{equation}
where
\begin{equation}
\hbar\omega^{\prime\prime}=\sqrt{2\hbar eB}v_F(1-\beta^2)^{3/4}
\end{equation}
and  $n=0,1,2,\cdots$. Note $n$ is positive and it means we are dealing with electrons, since holes ($n<0$) is out of the scope of the present work. Above,
\begin{equation}\label{beta}
\beta=\frac{E}{v_FB}<1
\end{equation}
and $k_y=2\pi l/L_y$ ($l=0,\pm1,\pm2$,...) is related to the size $L_y$ of the graphene along the $y$ direction. Note, increasing $\beta$ the Landau levels spacing decrease; and for a critical value ($\beta=1$), the entire Landau strucuture collapses\cite{prl_referee}. From this energy spectra, the \emph{grand} canonic potential per graphene area can be evaluated\cite{livro_grafeno,PLA_375_2011_3624,JPCM_22_2010_115302,PhysRevB.69.075104,PhysRevLett.103.025301,PhysRevLett.99.226803}:
\begin{align}
\Phi(T,&B,E)=\\\nonumber&\Phi_0(T)+\Phi_{no}(B,E)+\Phi_{per}(T,B,E)+\Phi_{cos}(T,B,E)
\end{align}
where $\Phi_0(T)$ is a contribution that depends only on the temperature; $\Phi_{no}(B,E)$ depends on both, magnetic and electric fields in a non-oscillatory fashion and does not depend on the temperature; $\Phi_{per}(T,B,E)$ has a periodic behavior and depends on the temperature, magnetic and electric fields and, finally, $\Phi_{cos}(T,B,E)$ has a cosine oscillatory character and also depends on the temperature, magnetic and electric fields. Thus, the entropy
\begin{equation}\label{defS}
S(T,B,E)=-\frac{\partial}{\partial T}\Phi(T,B,E)
\end{equation}
of the system can be obtained and it is easy to see that $S_{no}=0$ and
\begin{equation}
S(T,B,E)=S_0(T)+S_{per}(T,B,E)+S_{cos}(T,B,E)
\end{equation}

As mentioned above, to characterize the magnetocaloric effect, the magnetic entropy change must be considered
\begin{equation}
\Delta S(T,\Delta B,E)=S(T,B,E)-S(T,0,E)
\end{equation}
However, from equation \ref{beta}, we verify that the magnetic field can not be zero, otherwise $\beta$ would be bigger than one. Thus, there is a minimum value for the magnetic field: $B_{min}=E/v_F$ and therefore $B\rightarrow B_{min}$ implies to $\beta\rightarrow1$. The magnetic entropy change reads then as:
\begin{align}\label{MCE_finall}
\Delta S(T,\Delta B,E)=&\left[S_{per}(T,B,E)-S_{per}(T,E/v_F,E)\right]+\\\nonumber&\left[S_{cos}(T,B,E)-S_{cos}(T,E/v_F,E)\right]
\end{align}
similar to what was find to a 3D non-relativistic material\cite{MCE_OSC_DS} and graphene without external electric field\cite{mce_grafeno}.

\begin{figure}
\center
\includegraphics[width=7cm]{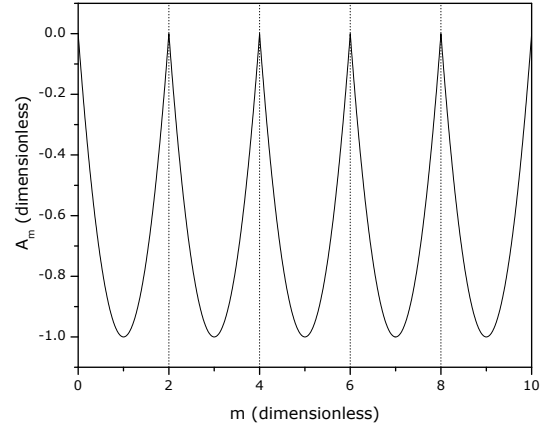}
\caption{Periodic function $A_m$, from equation \ref{am}.\label{Am}}
\end{figure}

To obtain the magnetic entropy change, only two \emph{grand} canonic potentials must be known: $\Phi_{cos}(T,B,E)$ and $\Phi_{per}(T,B,E)$. The first one, per graphene area, is\cite{JPCM_22_2010_115302,PLA_375_2011_3624}:
\begin{equation}\label{pqpp}
\Phi_{cos}(T,B,E)=\frac{(eB)^2v_F^2}{\pi\mu_0}\left(1-\beta^2\right)^{3/4}\sum_{k=1}^\infty\frac{\cos(\pi km)}{(\pi k)^2}\frac{x_k}{\sinh(x_k)}
\end{equation}
where $\mu_0=\hbar v_F\sqrt{N_0\pi}$ is the zero temperature and zero magnetic field chemical potential, i.e., the Fermi energy; and $N_0=10^{16}$ m$^{-2}$ is the density of charge carriers\cite{NN,JAP}. Above, $k$ is only the index of the summation and has no physical meaning. In addition:
\begin{equation}\label{mmesmo}
m=N_0\frac{\phi_0}{B}
\end{equation}
where $\phi_0=\pi\hbar/e=2.06\times10^{-15}$ Tm$^2$ is the magnetic flux quantum. Finally:
\begin{equation}\label{xk}
x_k=k\frac{\phi_0}{B}\frac{k_BT}{\tilde{v}_F}\frac{1}{(1-\beta^2)^{3/4}}
\end{equation}
where $\tilde{v}_F=\hbar v_F/2\pi\sqrt{N_0\pi}=9.43\times10^{-38}$Jm$^2$. Note both, $m$ and $x_k$ are dimensionless. The other \emph{grand} canonic potential of interest is\cite{JPCM_22_2010_115302,PLA_375_2011_3624}:
\begin{equation}\label{sth_phi}
\Phi_{per}(T,B,E)=\pi L_x\hbar v_F\beta \frac{N_0^2}{m^2}A_m\sum_{k=1}^\infty\frac{x_k}{\sinh(x_k)}
\end{equation}
where the periodic behavior lies on
\begin{equation}\label{am}
A_m=\left[m^2-2m(2\sigma+1)+4\sigma(\sigma+1)\right]
\end{equation}
Above, $\sigma=[m/2]$, where the $[\;]$ brackets mean the integer part of the argument. The periodic behavior of $A_m$ can be seen in figure \ref{Am}; and it is important to note this function is $-1$ of odd values of $m$ and, on the other hand, it assumes $0$ for even values of $m$.
Still considering equation \ref{sth_phi}, note it depends on $L_x$, the size of the graphene along $x$ direction. We considered $L_x=10^{-8}$ m\cite{PLA_375_2011_3624}. Those \emph{grand} canonic potentials (equations \ref{pqpp} and \ref{sth_phi}), obtained from references \onlinecite{JPCM_22_2010_115302,PLA_375_2011_3624}, consider all electronic levels of the system.

Thus, from the \emph{grand} canonic potentials above described, the corresponding entropies can be obtained:
\begin{equation}
S_{cos}(T,B,E)=2\pi k_B\frac{N_0}{m}\left(1-\beta^2\right)^{3/4}\sum_{k=1}^\infty\frac{\cos(\pi km)}{\pi k}\mathcal{T}(x_k)
\end{equation}
and
\begin{equation}
S_{per}(T,B,E)=2\pi^2 k_BL_x\beta A_m\frac{N_0}{m}\sqrt{N_0\pi}\sum_{k=1}^\infty k\mathcal{T}(x_k)
\end{equation}
Above
\begin{equation}\label{tau}
\mathcal{T}(z)=\frac{zL(z)}{\sinh(z)}\;\;\;\;\;\;\textnormal{where}\;\;\;\;\;\;L(z)=\coth(z)-\frac{1}{z}
\end{equation}
is the Langevin function. As usually considered for thermodynamic quantities with this kind of oscillations
\cite{greiner,MCE_OSC_DS,mce_grafeno}, we will consider only $k=1$ term: the hyperbolic sine function at the denominator of equation \ref{tau} rapidly damping the summation. Of course, it is an approximation; however, after this assumption, we can understand the physics behind this model. Thus, the relevant contributions to the entropy reads as:
\begin{equation}\label{cos}
S_{cos}(T,B,E)=2 k_B\frac{N_0}{m}\left(1-\beta^2\right)^{3/4}\cos(\pi m)\mathcal{T}(x)
\end{equation}
and
\begin{equation}\label{sth}
S_{per}(T,B,E)=2\pi^2 k_BL_x\beta A_m\frac{N_0}{m}\sqrt{N_0\pi}\mathcal{T}(x)
\end{equation}
where $x=x_1$.

To evaluate equation \ref{MCE_finall}, the $S(T,E/v_F,E)$ terms must be considered. From equation \ref{beta}, $B\rightarrow B_{min}=E/v_F$ implies to $\beta\rightarrow1$ and therefore $x\rightarrow\infty$ (see equation \ref{xk}). However, it is easy to see that $\mathcal{T}(x\rightarrow\infty)\rightarrow0$; and therefore $S(T,E/v_F,E)$ terms also tends to zero. Thus, the magnetic entropy change per graphene area reads as:
\begin{equation}\label{DSFIM}
\Delta S(T,\Delta B,E)=S_{per}(T,B,E)+S_{cos}(T,B,E)
\end{equation}
where the above contributions are in equations \ref{cos} and \ref{sth}.

\begin{figure}[b]
\center
\includegraphics[width=7cm]{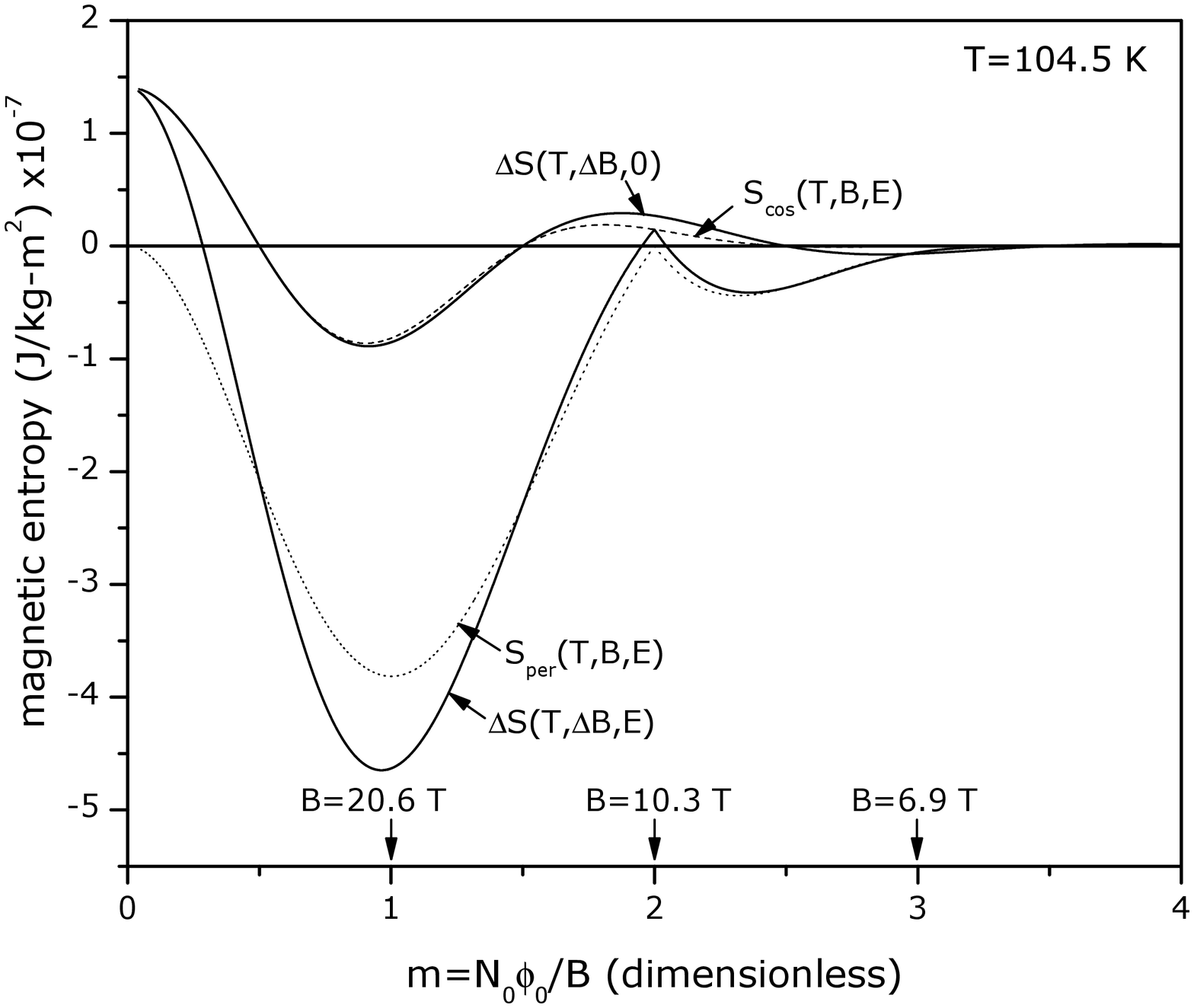}
\caption{Magnetic entropy change with applied electric field $\Delta S(T,\Delta B,E)$ (equation \ref{DSFIM}) and its contributions $S_{osc}(T,B,E)$ and $S_{per}(T,B,E)$. For the sake of comparison, also included is the case without applied field $\Delta S(T,\Delta B,0)$. These entropies are presented as a function of $m$, inversely proportional to the magnetic field (see equation \ref{mmesmo}). \label{DSn_MCE}}
\end{figure}

Considering first the case without electric field, i.e., $\beta=0$ (discussed in reference \onlinecite{mce_grafeno}), $S_{per}(T,B,E)$ term vanishes and therefore $S_{cos}(T,B,E)$ term promotes oscillations on $m$. It is represented in figure \ref{DSn_MCE} as $\Delta S(T,\Delta B,0)$. To understand the influence of the electric field on the magnetocaloric properties of graphenes, $\beta$ must be close to 1/2 (i.e.  $2E\approx v_FB$); otherwise, either $S_{cos}(T,B,E)\gg S_{per}(T,B,E)$ or $S_{cos}(T,B,E)\ll S_{per}(T,B,E)$. Note, from the case without electric field, the oscillations rapidly damping and for $m\gtrsim4$ there are no longer significant oscillations. In this direction, $m=4\rightarrow B= 5.15 $ T and therefore, due to the huge value of $v_F$, the electric field that allow a remarkable effect is of the order of $10^6$ V/m. Thus, figure \ref{DSn_MCE} presents both contributions, $S_{per}(T,B,E)$ and $S_{osc}(T,B,E)$, as well as the magnetic entropy change $\Delta S(T,\Delta B,E)$, considering $E=5\times10^6$ V/m. Note the magnetic entropy change with an applied electric field $\Delta S(T,\Delta B,E)$ is quite different from the case without electric field $\Delta S(T,\Delta B,0)$. The reason to choose 104.5 K will be clear further in the text. We emphasize this temperature is quite higher in comparison to the case of a non-relativistic 3D diamagnetic material (c.a. 1 K)\cite{mce_grafeno}.

\begin{figure}[t]
\center
\subfigure[Magnetic entropy change $\Delta S(T,\Delta B,E)$ and its contributions $S_{osc}(T,B,E)$ and $S_{per}(T,B,E)$, for $m=1$.]{\includegraphics[width=6cm]{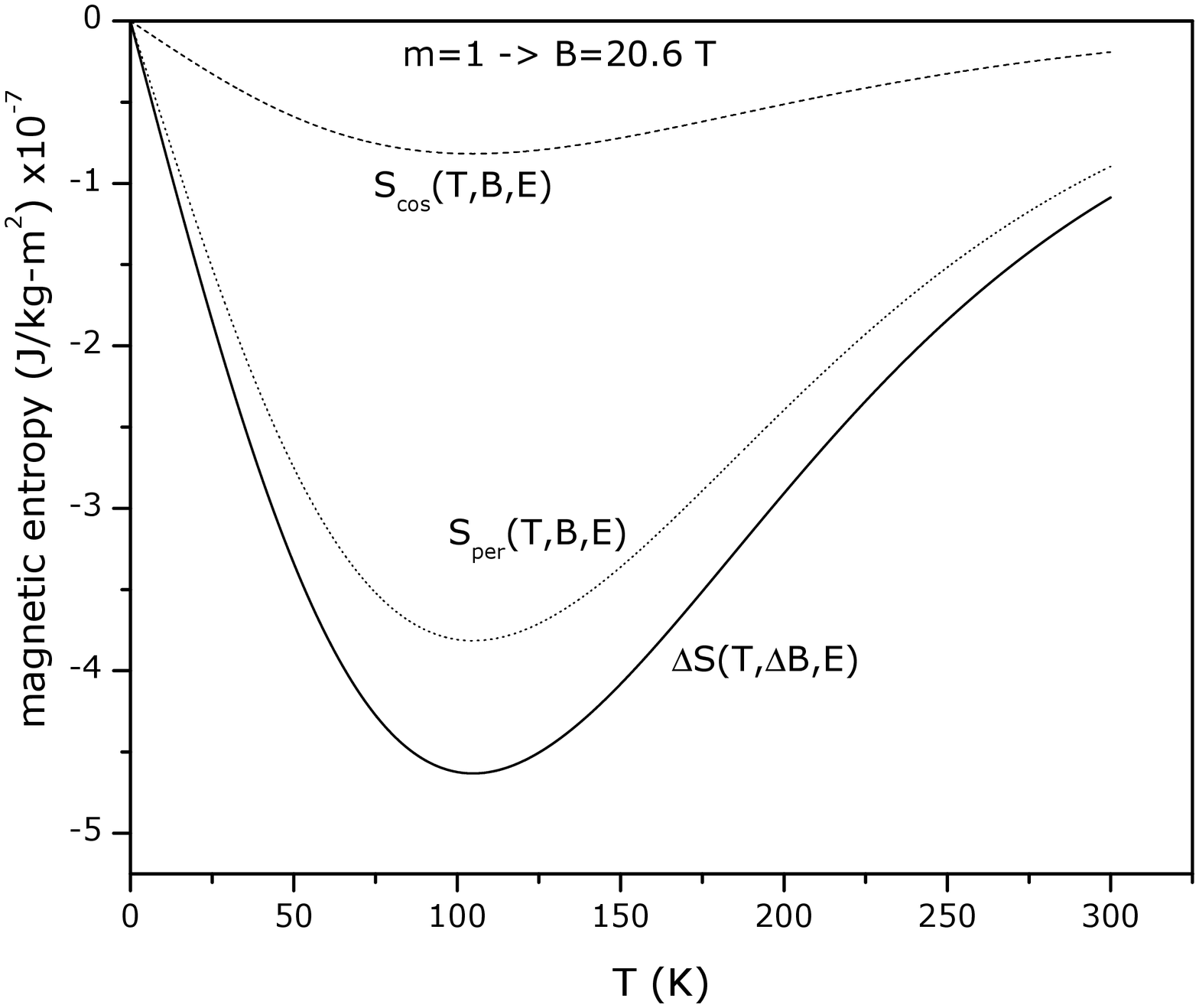}}
\subfigure[Magnetic entropy change $\Delta S(T,\Delta B,E)$ and its contributions $S_{osc}(T,B,E)$ and $S_{per}(T,B,E)$, for $m=2$.]{\includegraphics[width=6cm]{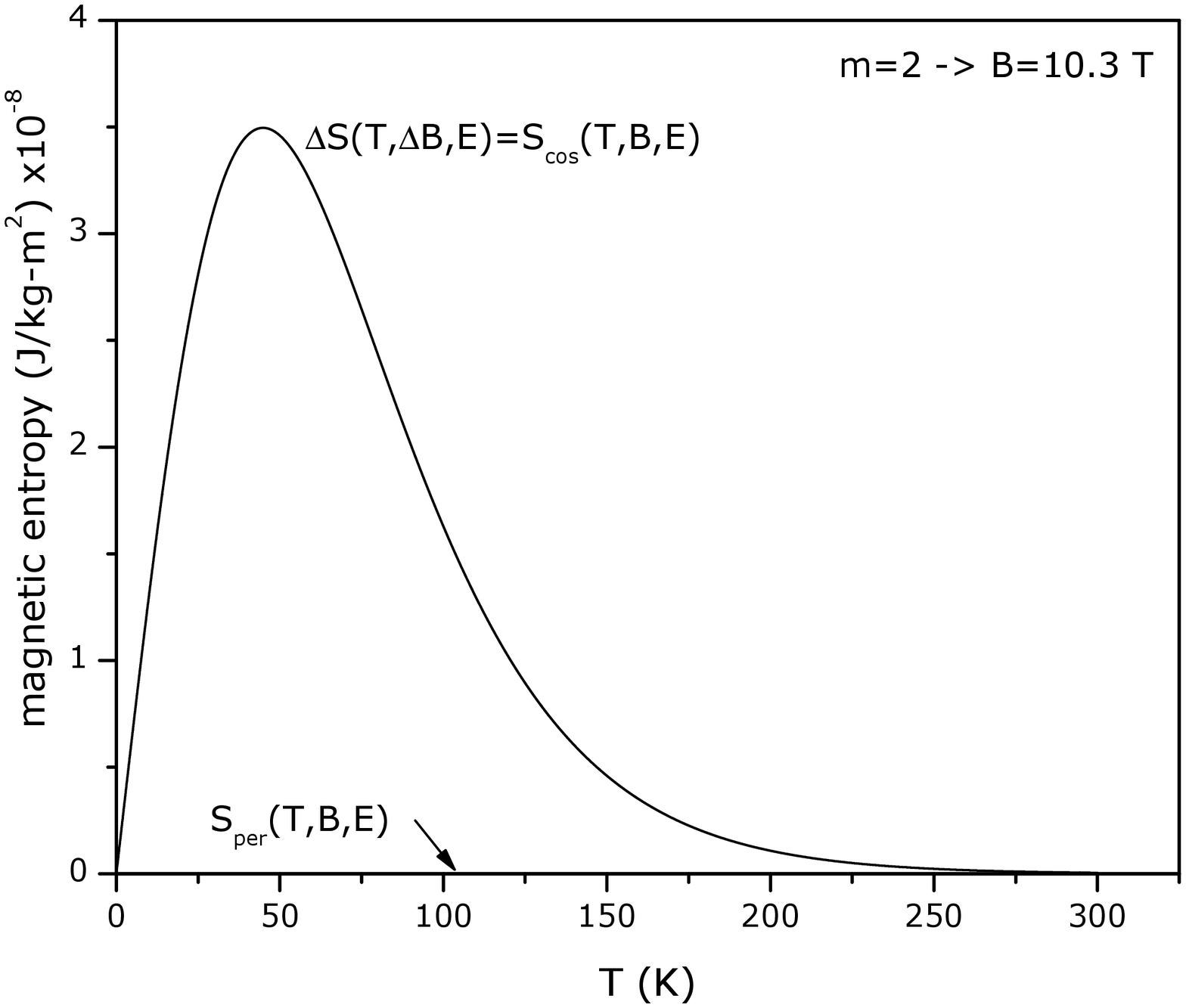}}
\subfigure[Comparison of $\Delta S(T,\Delta B,E)$ and $\Delta S(T,\Delta B,0)$, for some values of $m$. It represents the influence of the electric field on the magnetocaloric effect of graphenes.]{\includegraphics[width=6cm]{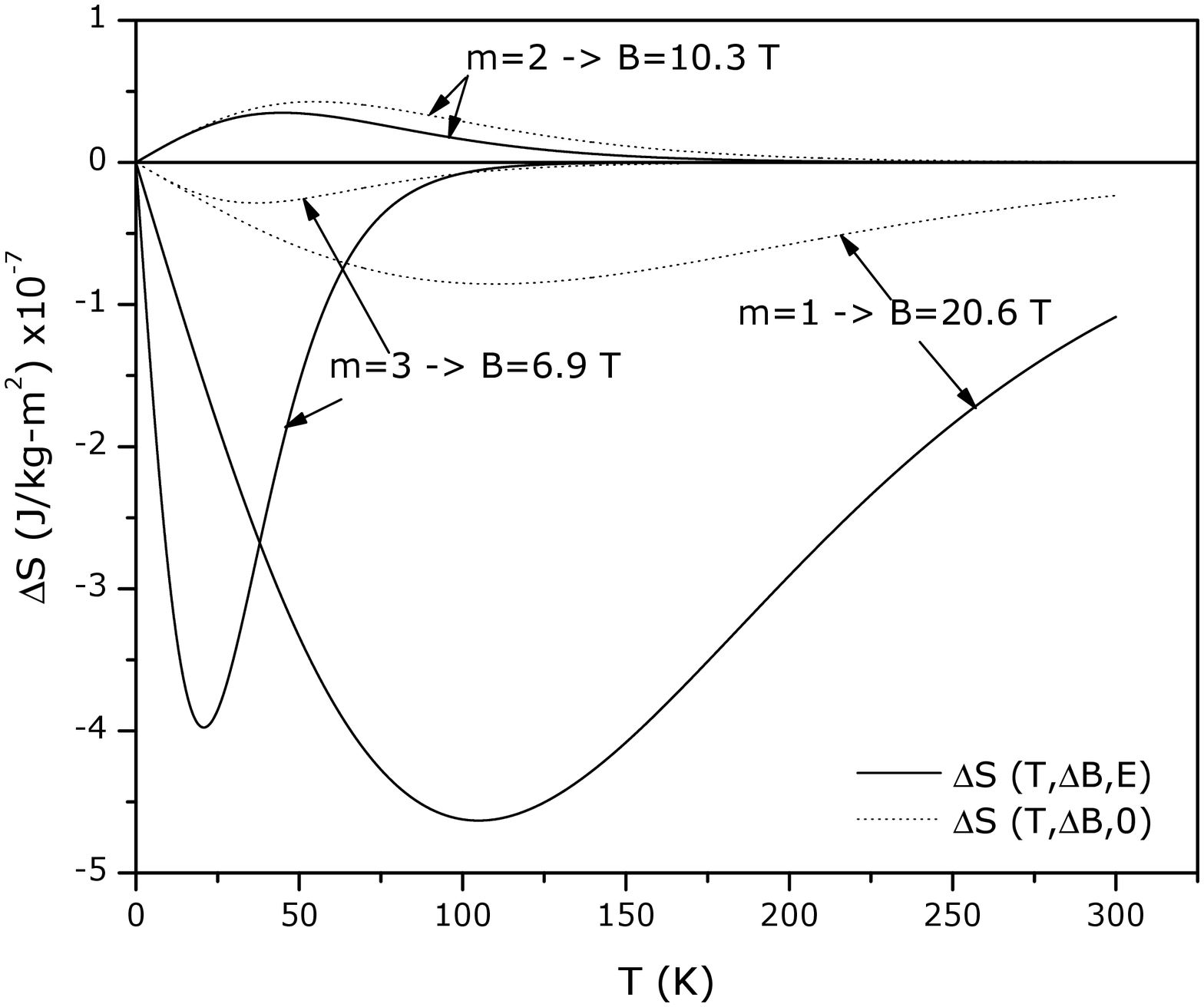}}
\caption{Temperature dependence of the magnetic entropy change of graphenes under a longitudinal applied electric field. \label{DST}}
\end{figure}

The influence of the electric field ($E=5\times10^6$ V/m), on the magnetic entropy change can also be seen as a function of temperature, as presented in figure \ref{DST}. The contributions $S_{osc}(T,B,E)$ and $S_{per}(T,B,E)$ to the magnetic entropy change $\Delta S(T,\Delta B,E)$ are shown in figures \ref{DST}(a) and (b), for $m=1$ and $m=2$, respectively. Note $S_{per}(T,B,E)=0$ for $m=2$, due to $A_m$. Figure \ref{DST}(c) shows the influence of the electric field on the magnetic entropy change, by comparing $\Delta S(T,\Delta B,E)$ and $\Delta S(T,\Delta B,0)$, for some values of $m$.

The behavior of the temperature dependence of the magnetic entropy change is rule by $\mathcal{T}(z)$ function (see equation \ref{tau}), that peaks at $z_{max}=1.6$ and therefore (from equation \ref{xk}):
\begin{equation}
T_{max}=1.6\frac{\tilde{v_F}}{k_B}\frac{N_0}{m}-1.6\frac{\tilde{v_F}}{k_B}\frac{N_0}{m}\frac{3\beta^2}{4}
\end{equation}
The first term only depends on the magnetic field and the second is quadratic on the electric field. Thus, an electric field decreases the temperature in which the magnetic entropy change is maximum (see figure \ref{DST}(c)). At this point, it is possible to understand the temperature chosen to figure \ref{DSn_MCE}, i.e., $T=T_{max}$ for $m=1$ and $E=5\times10^6$ V/m. 

On the other hand, an electric field changes the maximum value of the magnetic entropy change (see figure \ref{DST}(c)). To quantify the changing on the magnetic entropy change due to an applied electric field, let us consider an enhancement factor:
\begin{align}
\eta&=\frac{\Delta S(T_{max},\Delta B,E)}{\Delta S(T_{max},\Delta B,0)}\\\nonumber
&=\frac{S_{cos}(T_{max},B,E)+S_{sth}(T_{max},B,E)}{S_{cos}(T_{max},B,0)}
\end{align}
For odd values of $m$: $A_m=\cos(m\pi)=-1$ and therefore
\begin{align}
\eta&\approx1+\pi^2L_x\sqrt{N_0\pi}\beta-\frac{3}{4}\beta^2\\\nonumber
&\approx 1+18.40\beta-0.75\beta^2
\end{align}
Considering $E=5\times10^6$ V/m and $B=20.6$ T ($m=1$), thus $\beta=0.2427$. These values lead to $\eta=5.42$, and means that an electric field increases the magnetic entropy change, since $\eta>1$ (see figure \ref{DST}(c)). On the other hand, for even values of $m$: $A_m=0$ and $\cos(m\pi)=1$. Thus
\begin{equation}
\eta\approx1-\frac{3}{4}\beta^2
\end{equation}
Considering $E=5\times10^6$ V/m and $B=10.3$ T ($m=2$), thus $\beta=0.4854$. Note these values lead to $\eta=0.82$ and therefore an electric field decreases the magnetic entropy change, since $\eta<1$ (see figure \ref{DST}(c)).

It is worth to note the magnetic entropy change for these results are not, as usual, for $\Delta B:0\rightarrow B$, since the model does not allow $B=0$. From equation \ref{beta}, $B_{min}=E/v_F$ and therefore $B_{min}=5$ T.

Summarizing, we have described the influence of an applied electric field on the magnetocaloric properties of a 2D massless Dirac system: a graphene. For the case without electric field, this system presents an oscillating magnetic entropy change, due to a cosine term depending on $m$, inversely proportional to the magnetic field. An electric field promotes an extra entropy term, with a periodic pattern also on $m$ (this term has a maximum value for odd values of $m$ and vanishes for even values of $m$). These two contributions give the magnetic entropy change, in which the maximum value either increases or decreases due to an applied electric field, depending on the value of $m$. Finally, the temperature in which the maximum magnetic entropy change occurs decreases due to an applied electric field. 

We acknowledge FAPERJ, CAPES, CNPq and PROPPI-UFF for financial support.

\bibliography{bib}

\end{document}